\documentclass[twocolumn,aps,prl,superscriptaddress]{revtex4}
\usepackage{graphicx}
\usepackage{bm}
\usepackage{multirow}
\begin{document}
\title{Upper critical field of ${\bm p}$-wave ferromagnetic superconductors with orthorhombic symmetry}
\author{Richard A Klemm}\email{klemm@physics.ucf.edu}
\affiliation{Department of Physics, University of Central Florida, Orlando, FL 32816-2385 USA}
\author {Christopher L{\"o}rscher}
\affiliation{Department of Physics, University of Central Florida, Orlando, FL 32816-2385 USA}
\author{Jingchuan Zhang}
\affiliation{Department of Physics, University of Science and Technology Beijing, Beijing 100083, China}
\author{Qiang Gu}
\affiliation{Department of Physics, University of Science and Technology Beijing, Beijing 100083, China}

\begin{abstract}
 We extended the Scharnberg-Klemm theory of $H_{c2}(T)$ in $p$-wave superconductors with  broken symmetry to cases of partially broken symmetry in an orthorhombic crystal, as is appropriate for  the more exotic ferromagnetic superconductor UCoGe in strong magnetic fields. For some partially broken symmetry cases, $H_{c2}(T)$ can mimic upward curvature in all three crystal axis directions,  and reasonably good fits to some of the UCoGe data  are obtained.
\end{abstract}
\maketitle
\section{Introduction}
There has long been an interest in the possibility of superconductivity with the paired electrons having an order parameter consisting of a triplet spin configuration and the corresponding odd orbital symmetry \cite{ScharnbergKlemm1980,ScharnbergKlemm1985,VolovikGorkov1985,UedaRice1985,Blount1985,Sauls1994,Machida1999,Mineev1999}.  The simplest odd orbital symmetry has the $p$-wave form \cite{ScharnbergKlemm1980}.  In a crystal with non-cubic structure, there can be a variety of different $p$-wave states \cite{ScharnbergKlemm1980,ScharnbergKlemm1985,VolovikGorkov1985,UedaRice1985,Blount1985}. Depending upon the temperature $T$,  magnetic field ${\bm H}$, and  pressure $P$, there can be phases corresponding to  different triplet spin states \cite{Sauls1994,Machida1999,Mineev1999}.  One of the easiest ways to characterize the $p$-wave states is by measurements of the $T$ dependence of the upper critical field $H_{c2}(T)$ \cite{ScharnbergKlemm1980,ScharnbergKlemm1985}.  However, when multiple phases are present in the same crystal, as in UPt$_3$, a proper analysis  requires a variety of experimental results \cite{Sauls1994,Machida1999}.

Recently, a new class of ferromagnetic superconductors has been of great interest.  Presently this class  consists of UGe$_2$ \cite{Saxena2000}, UIr \cite{Akazawa2004}, URhGe \cite{Aoki2001}, and UCoGe \cite{Huy2007}, which except for UIr have orthorhombic crystal structures.  For  URhGe, the superconductivity arises within the ferromagnetic phase.  That is also true for  UCoGe at ambient pressure, but when sufficient pressure is applied, the ferromagnetic phase appears to disappear, leaving the superconducting phase without any obvious additional ferromagnetism \cite{Hassinger2008,Slooten2009}.  In the cases of UGe$_2$ and UIr, applying pressure within the ferromagnetic phase induces the superconductivity \cite{Saxena2000,Akazawa2004}.  In addition, polarized neutron studies have been interpreted as providing evidence for a field-induced ferrimagnetic state in UCoGe, with local moments of different magnitudes in opposite directions on the U and Co sites  \cite{Prokes2010}.  For a ferromagnetic superconductor with orthorhombic symmetry, the possible order parameter symmetries  were given by Mineev \cite{Mineev2002}.

 Hardy and Huxley measured ${\bm H}_{c2}(T)$ of URhGe at ambient pressure  in all three-crystal axis directions \cite{HardyHuxley2005}.  Using only one fitting parameter for each field direction,  they found that the Scharnberg-Klemm theory fit their data quantitatively \cite{HardyHuxley2005}, assuming the polar state with completely broken symmetry (CBS)\cite{ScharnbergKlemm1985}.    This remarkable fit for the low-field regime of the superconducting state in URhGe did not require any inclusion of the ferromagnetism into the theory, as the only apparent effect of the ferromagnetism was to give rise to a demagnetization effect jump in $H_{c2}$ at the superconducting transition temperature $T_c$.  In addition, $H_{c2}(0)$ exceeded the Pauli limit for all field directions measured, providing strong evidence of a parallel-spin pair state.

Upon the discovery of magnetic-field induced reentrant superconductivity in URhGe \cite{Levy2005}, much interest turned to the possible source of the high-field superconducting phase.  Then, superconductivity was discovered in UCoGe \cite{Huy2007}, and $H_{c2}(T)$ was measured for all three crystal axis directions \cite{Huy2008}, and all of the curves exhibited upward curvature  unrelated to dimensional-crossover effects \cite{KLB1975}.   Subsequently, a highly anomalous $S$-shaped $H_{c2}(T)$ curve was observed for $T<0.65T_c$ with ${\bm H}||\hat{\bm b}$ \cite{Aoki2009}.  Since ${\bm M}||\hat{\bm c}$ at low fields, this change in the ${\bm M}$  direction only  occurred in very pure, well-aligned samples.  This behavior may also have something to do with a reentrant phase, one that is close in field strength to the low-field phase \cite{Gasparini2010}

   The first attempts to describe  upward $H_{c2}(T)$ curvature in all crystal axis directions were based either upon ferromagnetic fluctuations \cite{Tada2011}, or upon     a crossover from one parallel-spin state to another \cite{Mineev2010}.  Meanwhile, a mean-field theory of the complementary effects of itinerant ferromagnetism and parallel-spin superconductivity was developed \cite{Nevidomskyy2005,Jian2010}.  To date, the field dependence of this mutual enhancement has not been investigated.  Here, we study the case  in which the $p$-wave pairing interaction strength is anisotropic, but finite in all  crystal directions.  Since ${\bm H}_{c2}$ is essentially isotropic in the $ab$ plane for samples of UCoGe with medium purity \cite{Huy2008}, we studied the partially broken symmetry (PBS) state as a function of the pairing interaction anisotropy.  This can give a kink in $H_{c2}(T)$ in at least  one field direction \cite{KlemmScharnberg1986}.
 \begin{figure}
 \center{\includegraphics[width=0.48\textwidth]{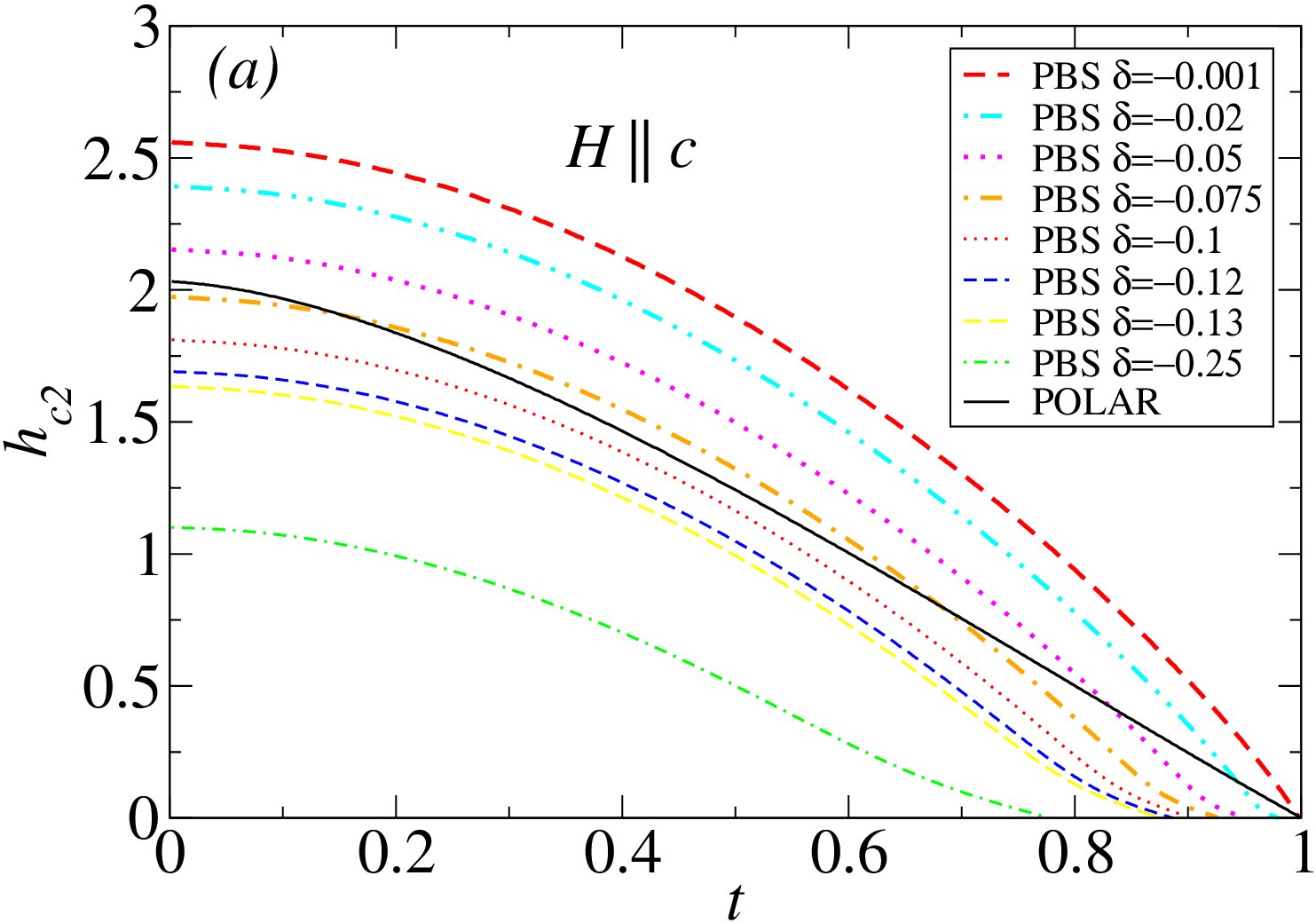}\vskip30pt
 \includegraphics[width=0.48\textwidth]{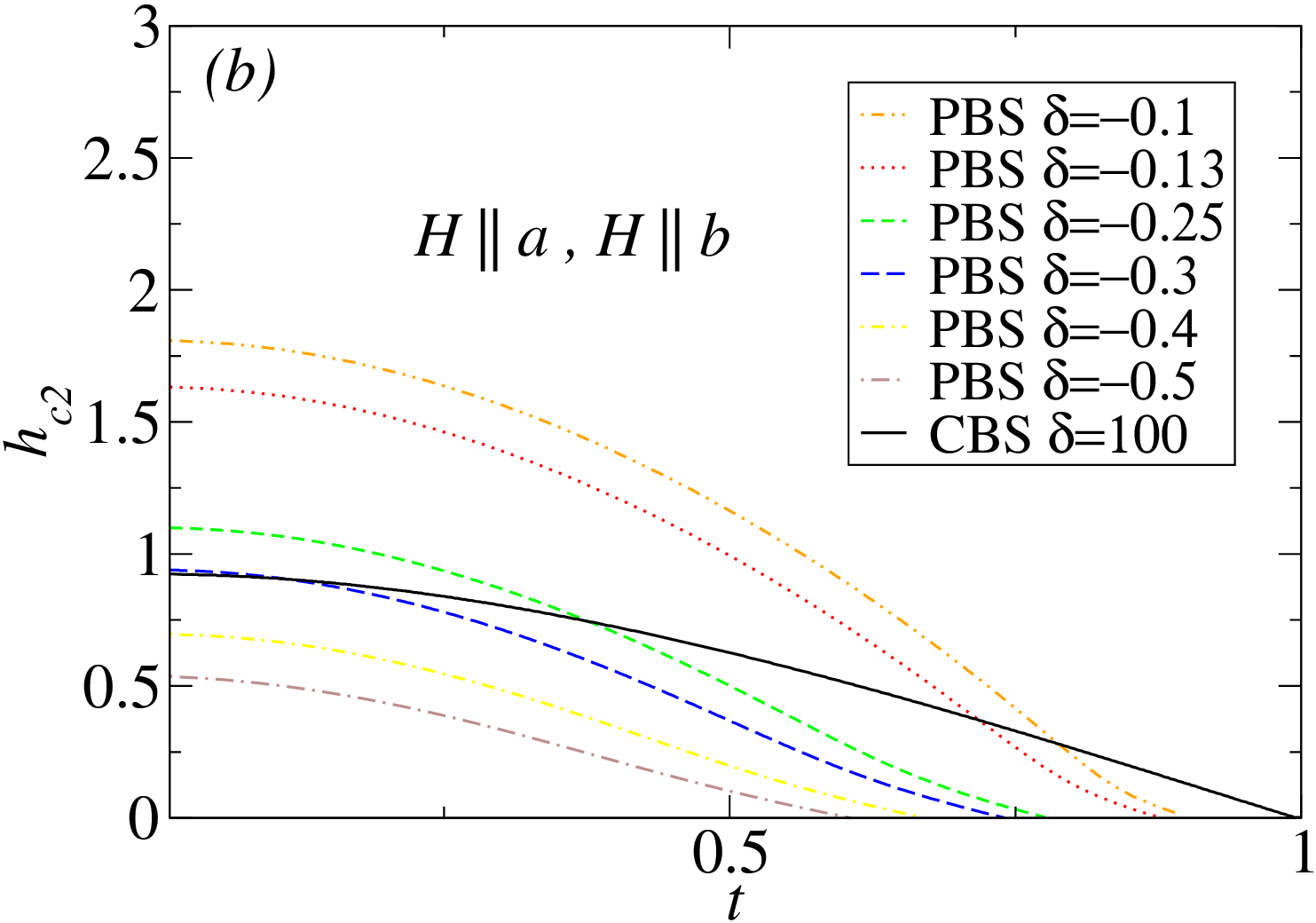}
 \caption{(a) Plots of $h_{c2,||c}=2eH_{c2}(m/m_{12})v_F^2/(2\pi T^c_c)^2$ versus $t=T/T_c^c$ for the polar state (solid black) and for a variety of PBSs with $-0.25\le\delta=\ln(T_c^{ab}/T_c^c)\le-0.001$.  (b) Plots of $h_{c2,\perp c}=2eH_{c2}(m/\sqrt{m_{12}m_3})v_F^2/(2\pi T^c_c)^2$ versus $t=T/T_c^c$ for the CBS state (solid black) and for various PBSs with  $-0.5\le\delta\le-0.1$.}}\label{fig1}
 \end{figure}

 \section{Upper critical field anisotropy of the PBS state}
 We assume a $p$-wave pairing interaction as in Eq. (1) of Ref. [2], where we take $V_3>V_2\ge V_1$.  We note that in the line following Eq. (7) of Ref. [2], $c_n=b_n^2/(\alpha_n^{+}\alpha_{n+2}^{+}-b_n^2)$ for $n\ge0$.  Then, for ${\bm H}||\hat{\bm e}_3$, the polar and two axial PBS states are obtained from
 \begin{eqnarray}
 \langle n|\Delta_{10}\rangle\alpha^{(p)}_{n}&=&0,\\
 (\langle n|\Delta_{11}\rangle\pm\langle n|\Delta_{1,-1}\rangle)\alpha_{n}^{\pm}&=&\mp b_{n-2}\langle n-2|\Delta_{11}\rangle\nonumber\\
 & &-b_n\langle n+2|\Delta_{1,-1}\rangle,
 \end{eqnarray}
 where $\alpha_n^{(p)}=[N(0)V_3]^{-1}-a_n^{(p)}$, $\alpha_n^{(-)}=[N(0)V_2]^{-1}-a_n^{(a)}$, $\alpha_n^{(+)}=[N(0)V_1]^{-1}-a_n^{(a)}$, and
 \begin{eqnarray}
 a_n^{(\lambda)}&=&\pi T\sum_{\omega_n}\int_0^{\pi}d\theta\sin\theta\Bigl(\begin{array}{c}3\cos^2\theta\\
 \frac{3}{2}\sin^2\theta\end{array}\Bigr)\int_0^{\infty}d\xi e^{-2\xi|\omega_n|}\nonumber\\
 & &\times e^{-\frac{1}{2}\zeta_{12}}L_n(\zeta_{12}),\\
 b_n&=&\pi T\sum_{\omega_n}\int_0^{\pi}d\theta\frac{3}{2}\sin^2\theta\int_0^{\infty}d\xi e^{-2\xi|\omega_n|}e^{-\frac{1}{2}\zeta_{12}}\nonumber\\
 & &\times F_n(\zeta_{12}),
 \end{eqnarray}
 where the upper (lower) terms in the parenthesis of $a_n^{(\lambda)}$ are for the polar ($\lambda=p$) and axial ($\lambda=a$) states, respectively, $\zeta_{12}=eH\xi^2v_F^2\sin^2\theta(m/m_{12})$, $m_{12}=\sqrt{m_1m_2}$, $m=(m_1m_2m_3)^{1/3}$, $L_n(z)$ are the Laguerre polynomials, $F_n(z)=\sum_{p=0}^n\frac{(-z)^{p+1}\sqrt{(n+2)(n+1)}n!}{p!(p+2)!(n-p)!}$, $N(0)$ is the single-spin density of states, and we set $\hbar=c=k_B=1$.  For the field along $\hat{\bm e}_1$ or $\hat{\bm e}_2$, one rotates the axes by $\pi/2$ about $\hat{\bm e}_2$ or $\hat{\bm e}_1$, respectively, and lets $m_{12}$ be replaced by $m_{23}$ or $m_{13}$, respectively.

 Since the low-field $H_{c2}(T)$ data of Huy {\it et al.} for UCoGe suggest that it has uniaxial symmetry, with ${\bm H}_{c2}||\hat{\bm a}\approx{\bm H}_{c2}||\hat{\bm b}$, in the following we will restrict our consideration to the $V_1=V_2$ case \cite{Huy2008}.  In order to fit the Aoki {\it et al.} data with the $S$-shaped $H_{c2,||b}(T)$ curve, it is necessary to use the full orthorhombic anisotropy in Eqs. (1)-(4), and to include the spontaneous and field-dependent magnetization.  To do so for the two axial states, one may obtain a recursion relation for either one of the amplitudes,  $\langle n|\Delta_{1,\pm1}\rangle$, by eliminating the other in Eq. (2), and then solving the recursion relation in terms of a continued fraction.  In Fig. 1(a), we plotted $h_{c2,||c}=2eH_{c2}(m/m_{12})v_F^2/(2\pi T^c_c)^2$ versus $t=T/T_c^c$ for the polar state and for a variety of PBSs  with $-0.25\le\delta<0$, where $\delta=\ln(T_c^{ab}/T_c^c)$.  Note that these PBS states all have upward curvature, but since $T_c^c>T_c^{ab}$, the polar state dominates near to $T_c^c$.  However, for $-0.05\le\delta<0$, there is a single kink in $H_{c2,||c}(T)$, and for $\delta=-0.075$, there are two kinks, due to two crossovers between the polar and PBS states.  For $\delta\le-0.1$, there is no crossover to a PBS state.  In Fig. 1(b), we plotted $h_{c2,\perp c}=2eH_{c2}(m/\sqrt{m_{12}m_3})v_F^2/(2\pi T_c^c)^2$ versus $t=T/T_c^c$ for the CBS state and for various PBSs with  $-0.5\le\delta<0$.  In this case, the CBS state dominates near to $T_c^c$, but there is a crossover to the PBS state for $-0.3\le\delta<0$, resulting in a single kink in $H_{c2,\perp c}(T)$.

 \section{Fits to the Huy {\it et al.} UCoGe $H_{c2}(T)$ data}

 As a starting point, to see if there is any possibility of fitting the least anomalous region of the $H_{c2}(T)$ curves obtained for UCoGe, we  assume uniaxial anisotropy and fit the data of Huy {\it et al.}\cite{Huy2008}.  In Fig. 2(a), the best fit to the ${\bm H}||\hat{\bm c}$ data are for $\delta=-0.065$ is shown.  In Fig. 2(b), the best fit to the ${\bm H}||\hat{\bm a}$ and ${\bm H}||\hat{\bm b}$ data are for $\delta=-0.185$, which shows a distinct crossover from the CBS to the PBS state.  We remark that when the spontaneous magnetization is along the c-axis direction, the fitting to the data in Fig. 2(a) would be altered.
 \begin{figure}
 \center{\includegraphics[width=0.4\textwidth]{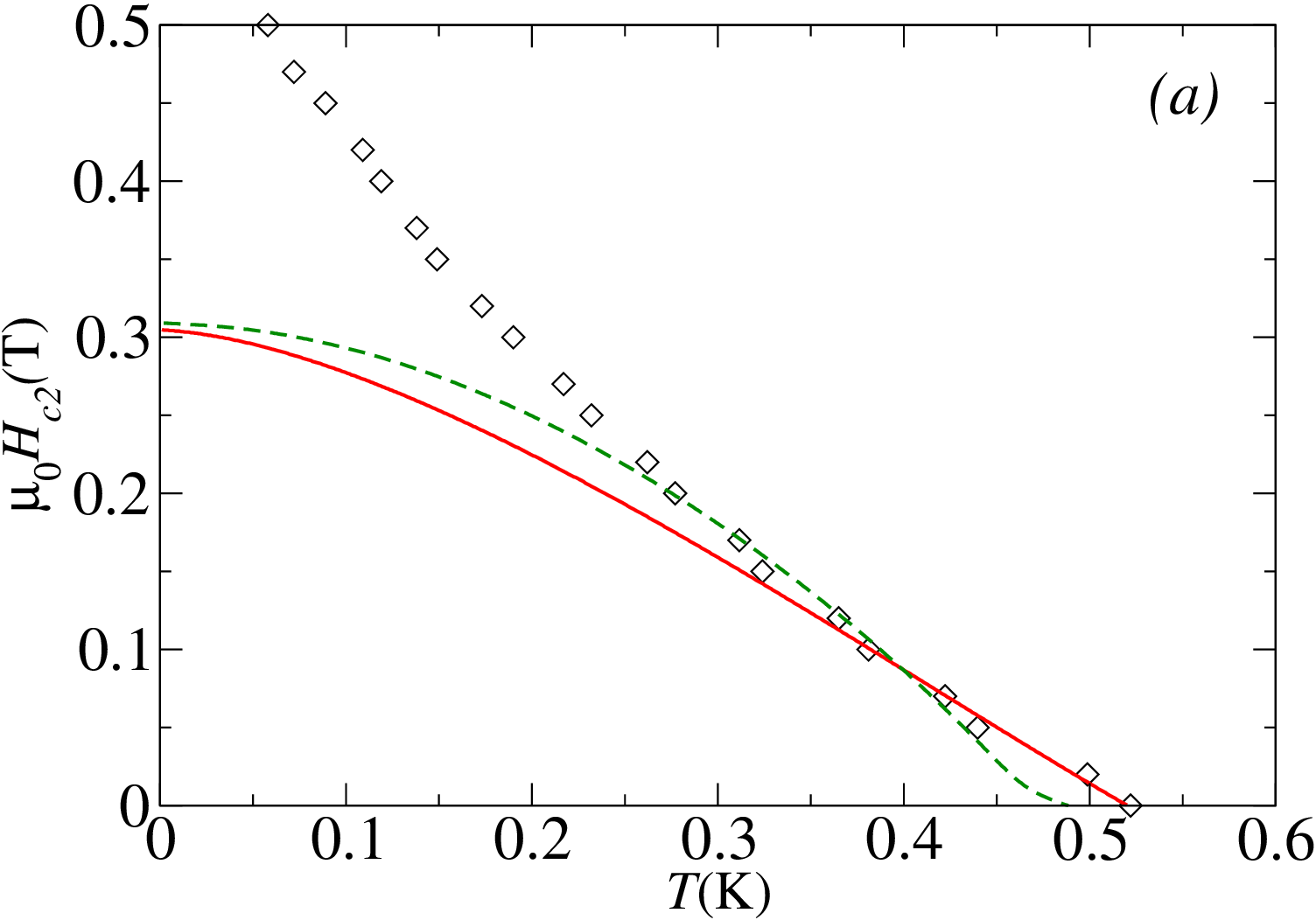}\vskip30pt
 \includegraphics[width=0.4\textwidth]{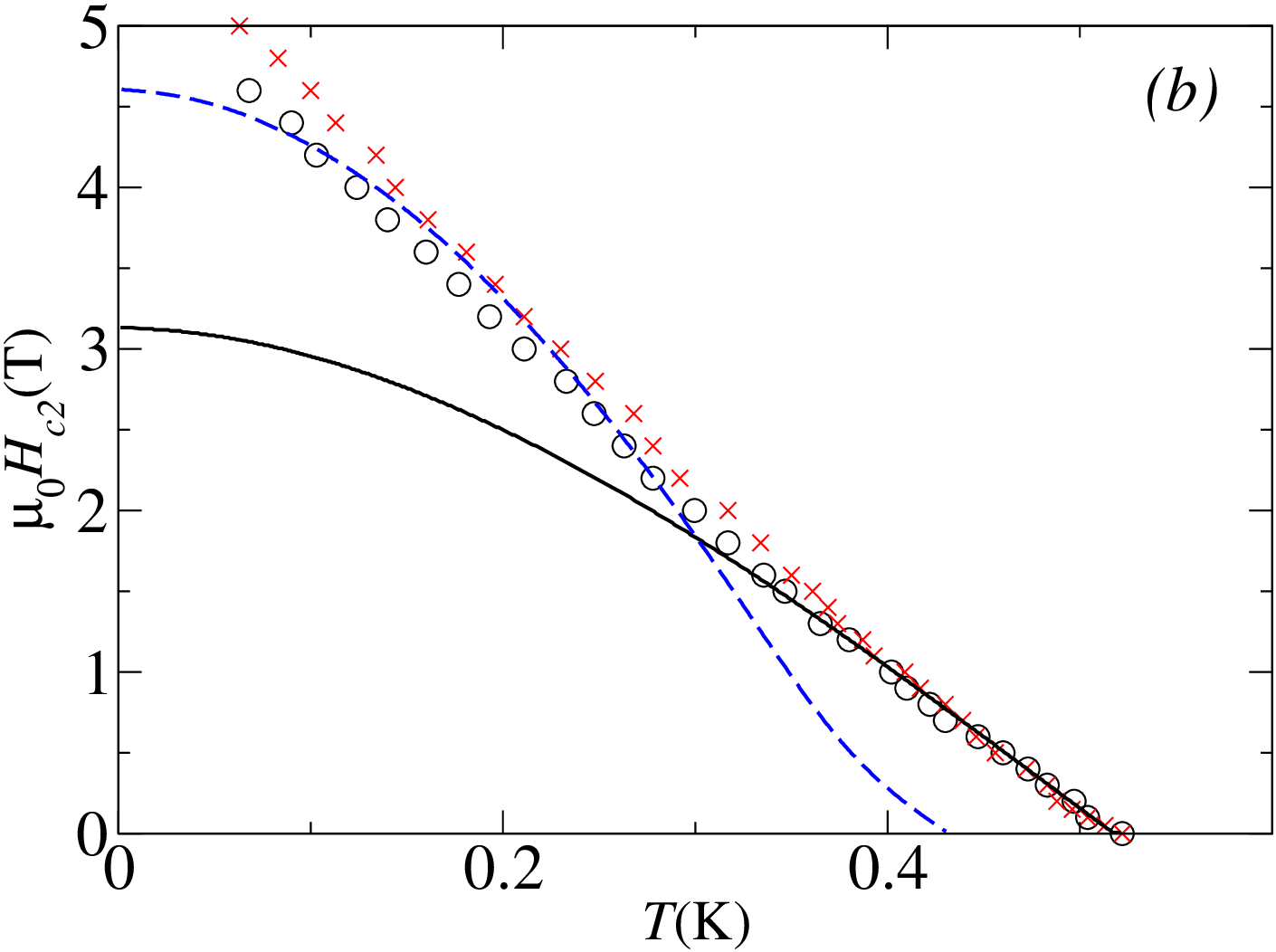}\caption{Best fits to the data  of Huy {\it et al.} for $\mu_0H_{c2}(T)$ in medium purity UCoGe \cite{Huy2008}. (a) ${\bm H}||\hat{\bm c}$.  Open black diamonds: data. The red solid and green dashed curves are for the polar state and the PBS states with $\delta=-0.065$, respectively.  (b) Data for ${\bm H}||\hat{\bm b}$ (red crosses) and ${\bm H}||\hat{\bm a}$ (open black circles).  The solid black and blue dashed curves are for the CBS state and the PBS state with $\delta=-0.185$, respectively.  The slopes at $T_c$ were adjusted to fit the data.}}\label{fig2}
 \end{figure}

 \section{Conclusions}
 We found that it is possible to fit the upward curvature of the $H_{c2}(T)$ data from medium-purity UCoGe using a crossover from the polar/CBS state to a PBS state.  However, without taking account of the spontaneous magnetization, the best fit values of $T_{c}^{ab}$ [or $\delta=\ln(T_c^{ab}/T_c^c)$] are different for different field directions.  At the very least, the spontaneous and field-dependent magnetization should be included in future fits, using an anisotropic intinerant ferromagnetic superconductor model similar to that previously studied \cite{Nevidomskyy2005,Jian2010}.

 \section{Acknowledgments}
 The authors are grateful to Prof. A. de Visser for providing the data of Huy {\it et al.}\cite{Huy2008}.  QG acknowledges the Specialized Research Fund for the Doctoral Program of Higher Education of China (no. 20100006110021).


\end{document}